\documentclass[12pt,a4paper]{conference}
\usepackage{fancyhdr}
\usepackage{graphicx,amsmath,amssymb,cite}
\usepackage{multind}
\DeclareMathSymbol{\varGamma}{\mathord}{letters}{"00}
\begin{document}
\newcommand{\vek}{\mbox{\boldmath${\rm k}$}}
\newcommand{\vp}{\varphi}
\newcommand{\vt}{{\vec \tau}}

\Chapter{The nature of the light scalar mesons from their radiative decays}
           {The nature of the light scalar mesons\ldots}{A. V. Nefediev}

\addcontentsline{toc}{chapter}{{\it A. Nefediev}} \label{authorStart}

\begin{raggedright}
{\it A. V. Nefediev}\\
Institute of Theoretical and Experimental Physics, Moscow, 
Russia\footnote{Institute of Theoretical and Experimental 
Physics, 117218, B.Cheremushkinskaya 25, Moscow, Russia.}\\
\bigskip\bigskip
\end{raggedright}

\begin{center}
\textbf{Abstract}
\end{center}
The nature of the light scalar mesons is one of the most intriguing open 
challenges in hadronic spectroscopy. It is argued that radiative decays 
involving these scalars can serve as an important decisive tool in 
establishing their nature. In particular, special emphasis is made on the 
radiative decays of the scalars themselves (in addition to the radiative 
decays of the $\phi$-meson with the scalars appearing in the final state), 
including their two-photon decays. All the above mentioned processes are 
considered in detail in the (point-like) kaon molecule model of the 
scalars and explicit predictions for the decay widths are made. In 
addition, finite-range corrections to the point-like results are 
investigated, with a special attention payed to gauge invariance of the 
decay amplitude. Finally, the conclusion is made that experimental data on 
the radiative decays with the light scalar mesons involved strongly 
support the molecule assignment for the latter.

\section{Introduction}

Understanding the properties of light scalar mesons is one of the most 
challenging problems of the hadrons spectroscopy. In particular, 
investigations of the nature of the $a_0(980)$ and $f_0(980)$ mesons 
attract considerable theoretical and experimental efforts. This interest 
should not come as a surprise since the given states reside at the very 
kaon--antikaon threshold and thus the admixture of the kaon molecular 
component in the wave function is expected to be large. Indeed, 
experimental data \cite{SND,CMD,KLOE} unambiguously show a prominent $K 
\bar K$ contribution. Other assignments for these mesons are also 
suggested and studied in the literature, such as the genuine $q\bar{q}$ 
assignment \cite{qq}, or the compact four--quark assignment 
\cite{Jaffe,4q}. Many experimental tests have been suggested so far in 
order to distinguish between these assignments and thus to disclose the 
nature of the scalars $a_0/f_0(980)$. For example, the importance of 
measurements of the radiative decays of the $\phi (1020)$ to scalar mesons 
was argued in \cite{AI}. In the meantime, another class of radiative 
decays --- the radiative decays of the scalars themselves --- can be 
studied as well and provide new important data. As compared to the 
radiative decays $\phi\to\gamma S$ ($S=a_0/f_0$), the decays $S\to\gamma 
V$ ($V=\rho,\omega,\gamma$) possess a number of advantages, such as a 
considerable phasespace in the final state and a possibility to probe the 
nonstrange component of the scalars w.f. These radiative decays can serve 
therefore as a complementary source of information and to deliver decisive 
information on the structure of these long--debated objects \cite{HKKN}.

Both types of radiative decays can be described with the single vertex 
function $VS\gamma$. Gauge invariance imposes quite restrictive constrains 
on the structure of the transition matrix element: 
$iW^{\mu\nu}=M(m_V^2,m_S^2)[P_V^\mu 
P_\gamma^\nu-g^{\mu\nu}(P_VP_\gamma)]$, where $P_V$ and $P_\gamma$ are the 
vector and the photon four-momenta.
 
\section{Evaluation of the decay widths in various assignments for the scalars} 
 
In the quark--antiquark assignment, the $a_0/f_0(980)$ mesons are treated 
as the genuine quark--antiquark ${}^3P_0$ states. Their radiative decays 
can be studied in the framework of nonrelativistic quark models 
\cite{KR,qgamgam} yielding the width of 125 keV, for the decays 
$a_0\to\gamma\omega$ and $f_0\to\gamma\rho$, 14 keV, for the decays 
$a_0\to\gamma\rho$ and $f_0\to\gamma\omega$, and 4.5 keV, for the decays 
$a_0/f_0\to\gamma\gamma$. The radiative decays widths of the genuine 
quark--antiquark mesons $f_1(1285)$ ($\Gamma(f_1(1285)) \to \gamma 
\rho)=1320 \pm 312 {\rm keV}$) and $f_2(1270)$ ($\Gamma(f_2(1270) \to 
\gamma \gamma)=2.61 \pm 0.30~{\rm keV}$) were used here in order to fix 
the radial w.f. matrix element \cite{HKKN}.
 
In the molecule assignment for the scalars, the radiative decays proceed 
via a kaon loop, and the scales involved into the problem possess the 
hierarchy $\varepsilon\ll m\lesssim\beta$, where $\beta$ is the intrinsic 
scale of the binding force, $m$ is the kaon mass, and $\varepsilon=2m-m_S$ 
is the binding energy. It was argued in \cite{2gam} that, for the 
realistic values of the parameters ($\beta\approx m_\rho\approx 800$ MeV, 
$m=495$ MeV, and $\varepsilon=10$ MeV), this hierarchy can be achieved 
starting prom the point-like limit of $\beta\to\infty$ and taking into 
account finite-range corrections in the inverse power of $\beta$. The 
point-like $SK\bar{K}$ coupling constant, 
$g_S^2/(4\pi)=32m\sqrt{m\varepsilon}\approx 1.12\;{\rm GeV}^2$ was 
obtained in \cite{mol0}. The two remaining couplings can be obtained from 
the $\rho\pi\pi$ constant ($g_V=g_\rho=g_\omega=\frac{1}{2}g_{\rho \pi 
\pi}\approx 2.13$) and from the total width of the $\phi$ 
($g_\phi^2/(4\pi)\approx 1.77$). Then the point-like predictions for the 
widths are $\Gamma(\phi\to\gamma S)=0.6$ keV, $\Gamma(S\to\gamma V)=3.4$ 
keV, and $\Gamma(S\to\gamma\gamma)=0.22$ keV. It can be demonstrated 
explicitly that no large corrections to these results, of order ${\cal 
O}(m^2/\beta^2)$, appear \cite{mol0,Markushin,2gam}. Thus one concludes 
that inclusion of the finite--range corrections does not change these 
prediction appreciably, giving only moderate (of order $10\div 20\%$ in 
the amplitude) corrections, provided they are included in a 
self-consistent and gauge-invariant way \cite{Markushin,mol0,2gam}.

\section{Conclusions}

In Table~1 we give the widths for the radiative decays involving scalars. 
Comparing the predictions made in the quark--antiquark and molecule 
assignment with the experimental data we conclude that the molecule 
picture is strongly supported by the data (Belle reports the new result 
$\Gamma(f_0\to\gamma\gamma)=0.205_{-0.83-0.117}^{+95+0.147}$ keV 
\cite{belle07} which is in even better agreement with the molecule 
prediction). An important property revealed by the radiative decays of the 
scalars is that the theoretical predictions for these decays differ 
drastically depending on the assignment made for the nature of the 
scalars. This makes such radiative decays an important tool in 
establishing the structure of the $a_0/f_0(980)$ mesons. We conclude that 
experimental data on the radiative decays $a_0/f_0\to\gamma\rho/\omega$ 
are strongly needed, as an important, and possibly decisive, source of 
information about the scalar mesons.

\begin{table}[t]
\caption{The widths (in keV) of the radiative decays involving scalars; 
$\theta$ is the (small) $\phi-\omega$ mixing angle.}
\begin{center}
\begin{tabular}{l|l|l|l}
\hline
\hline
&Quark--antiquark&Molecule&Data (PDG)\\
\hline
$\phi\to\gamma a_0$&$0.37\sin^2\theta$&0.6&$0.32\pm0.02$\\
\hline
$\phi\to\gamma f_0(\bar{n}n)/f_0(\bar{s}s)$&$0.04\sin^2\theta/0.18$&0.6&$0.47\pm0.03$\\
\hline
$a_0\to\gamma\gamma$&$2\div 5$&0.22&$0.30\pm 0.10$\\
\hline
$f_0\to\gamma\gamma$&$2\div 5$&0.22&$0.29^{+0.07}_{-0.09}$\\
\hline
$a_0\gamma\omega/\rho$&125/14&3.4&\\
\cline{1-3}
$f_0(\bar{n}n)\gamma\rho/\omega$&125/14&3.4&pending\\
\cline{1-3}
$f_0(\bar{s}s)\gamma\rho/\omega$&$0/31\sin^2\theta$&3.4&\\
\hline
\hline
\end{tabular}
\end{center}
\end{table}
\section*{Acknowledgments}

This work was supported by the Federal Agency for Atomic Energy of Russian 
Federation and by grants NSh-843.2006.2, DFG-436 RUS 113/820/0-1(R), 
RFFI-05-02-04012-NNIOa, and PTDC/FIS/70843/2006-Fi\-si\-ca.

\end{document}